\documentclass[journal,twoside,web]{IEEEtran}

\usepackage{cite}
\usepackage{amsmath,amssymb,amsfonts}
\usepackage{algorithmicx}
\usepackage{graphicx}
\usepackage{textcomp}
\usepackage{multirow}
\usepackage{comment}
\usepackage{url}
\usepackage{soul}
\usepackage{xcolor}
\usepackage{algorithm,algpseudocode}
\makeatletter
\newcommand{\linethrough}{\mathpalette\@thickbar}
\newcommand{\@thickbar}[2]{{#1\mkern0mu\vbox{
    \sbox\z@{$#1#2\mkern-1.5mu$}%
    \dimen@=\dimexpr\ht\tw@-\ht\z@+2\p@\relax 
    \hrule\@height0.5\p@ 
    \vskip\dimen@
    \box\z@}}
}
\makeatother

\def\BibTeX{{\rm B\kern-.05em{\sc i\kern-.025em b}\kern-.08em
T\kern-.1667em\lower.7ex\hbox{E}\kern-.125emX}}
    
\begin{document}
\title{Boundary Constraint-free Biomechanical Model-Based Surface Matching for Intraoperative Liver Deformation Correction}
\author{Zixin Yang, Richard Simon, Kelly Merrell, Cristian. A. Linte.
    \thanks{This work was supported by the National Institutes of Health - National Institute of General Medical Sciences under Award No. R35GM128877 and the National Science Foundation - Office of Advanced Cyber-infrastructure under Award No.1808530 and the Division of Chemical, Bioengineering and Transport Systems under Award No. 2245152. (Corresponding author: Zixin Yang.)}
    \thanks{Zixin Yang, Richard Simon, Kelly Merrell, and Cristian. A. Linte are with the Center for Imaging Science and Department of Biomedical Engineering, Rochester Institute of Technology, Rochester, NY14623 USA. (Email:{yy8898, rasbme, kam1217, calbme@rit.edu})
    }
    \thanks{Zixin Yang and Richard Simon contributed equally.}
    } 
   
    \maketitle

\maketitle

\begin{abstract} 

In image-guided liver surgery, 3D-3D non-rigid registration methods play a crucial role in estimating the mapping between the preoperative model and the intraoperative surface represented as point clouds, addressing the challenge of tissue deformation. Typically, these methods incorporate a biomechanical model, represented as a finite element model (FEM), \textcolor{black}{into the strain energy term}  to regularize a surface matching term. 
\textcolor{black}{We propose a 3D-3D non-rigid registration method that incorporates a modified FEM into the surface matching term. The modified FEM alleviates the need to specify boundary conditions, which is achieved by modifying the stiffness matrix of a FEM and using diagonal loading for stabilization. As a result, the modified surface matching term does not require the specification of boundary conditions or an additional strain energy term to regularize the surface matching term.}
Optimization is achieved through an accelerated gradient algorithm, further enhanced by our proposed method for determining the optimal step size. We evaluated our method and compared it to several state-of-the-art methods across various datasets. Our straightforward and effective approach consistently outperformed or achieved comparable performance to the state-of-the-art methods. Our code and datasets are available at \url{https://github.com/zixinyang9109/BCF-FEM}.
\end{abstract}


\section{Introduction}
\label{sec:introduction}

\IEEEPARstart{I}{n} image-guided liver surgery (IGLS), preoperative imaging modalities such as Computed Tomography (CT) or Magnetic Resonance Imaging (MRI) provide detailed insights into the liver's internal structure, revealing information about tumors and vessels. However, these modalities are not commonly utilized intraoperatively due to practical constraints. Instead, alternative imaging techniques are employed during surgery, including tracked ultrasound \cite{heiselman2020intraoperative, smit2023ultrasound}, cone-beam CT (CBCT) \cite{peterlik2018fast}, and optical methods \cite{rucker2013mechanics, suwelack2014physics, modrzejewski2019vivo, mendizabal2023intraoperative,disparity}.

To integrate subsurface information from preoperative CT/MRI into the surgeon's intraoperative view, registration methods are employed to map critical information from the preoperative images into the intraoperative scene. Rigid registration is often used to align data collected in different coordinates. However, the initial configuration of the liver at the start of the intervention can significantly deviate from its preoperative state. Studies indicate that the deformation of the anterior surface of the liver can exceed 10 mm during laparoscopic liver surgery (LLS) and 7 mm during open liver surgery (OLS) \cite{heiselman2018characterization}. In clinical contexts, the target registration error is expected to be below 5 mm \cite{rucker2013mechanics}. Therefore, non-rigid registration may become essential to correct for such deformations.

Registration methods are broadly categorized into intensity-based \cite{lee2010non} and geometric-based \cite{labrunie2023automatic, espinel2021using,collins2020augmented} approaches based on the type of information utilized to establish correspondences between the pre- and intraoperative data. Intensity-based methods employ pixel or voxel intensities for aligning images. However, challenges arise from intensity variations across different imaging modalities, such as CT to ultrasound, impacting the efficacy of these methods.



Geometric-based registration methods can be further classified based on the availability of 3D information extracted from intraoperative imaging. For scenarios where only 2D information is available, as with intraoperative monocular laparoscopes, 3D-2D registration methods are typically employed \cite{labrunie2023automatic, espinel2021using}. In contrast, when intraoperative imaging modalities provide 3D information, usually presented as point clouds, 3D-3D registration methods \cite{heiselman2020intraoperative,modrzejewski2019vivo, suwelack2014physics} are preferred, as including 3D information introduces additional constraints to enhance registration accuracy \cite{collins2020augmented}. 


This study focuses on the 3D-3D non-rigid registration problem in IGLS. Unlike the typical 3D-3D non-rigid registration problem in the computer vision  \cite{deng2022survey}, where the primary focus is achieving surface matching accuracy, the problem in IGLS emphasizes the necessity for the estimated organ deformation to be realistic, with a crucial consideration for the deformation beneath the tissue surface. Biomechanical models, specifically represented as finite element models (FEMs), have been extensively employed in modeling tissue deformation \cite{zhang2017deformable} and are key to solving the 3D-3D non-rigid registration problem \cite{rucker2013mechanics, mestdagh2022optimal, suwelack2014physics}. The biomechanical model not only constraints displacements to enforce realistic deformations but also provides insights into volumetric deformation, capturing the complexities of organ deformations beyond surface-level changes.



\subsection{Related work}




We focus on registration techniques that utilize a FEM-based biomechanical model to align a preoperative model with an intraoperative point cloud acquired during the surgical procedure. This alignment subsequently aids in determining the locations of subsurface anatomical structures within the intraoperatively deformed liver.






In energy minimization approaches, it is common to integrate a FEM as a strain/deformation energy term to regularize a data term that ensures surface alignment between the preoperative model and the observed intraoperative point cloud data. The strain energy arises within the preoperative model due to the nonrigid displacement field. Rucker \textit{et al.} \cite{rucker2013mechanics} assumed the posterior side of the organ drove the deformation, then iteratively updated parameters that described the boundary condition and rigid transformation.  Based on the similar parameters formation, the framework was further extended to impose constraints from the ligament \cite{heiselman2018characterization} of the liver and vascular features \cite{heiselman2020intraoperative}. The method proposed by Modrzejewski \textit{et al.} \cite{modrzejewski2019vivo} includes an iterative closest points (ICP)-based data term, a collision term preventing self-collision, and a strain energy term. Khallaghi  \textit{et al.} \cite{khallaghi2015biomechanically} proposed a framework referred to as the GMM-FEM method in the context of MRI and transrectal ultrasound (TRUS) prostate registration. This method incorporates the FEM into a coherent point drift (CPD) algorithm \cite{myronenko2010point} as the regularization term and uses a Gaussian-mixture model to represent the preoperative surface. The strain energy term can also be minimized without the data term, as shown in the work presented by Peterlik \textit{et al.} \cite{peterlik2018fast}, where strain energy is minimized while satisfying the geometric constraints searched from ICP.

A limited number of methods directly incorporate the FEM into the data term. Mestdagh et al. \cite{mestdagh2022optimal, Mestdagh2022} integrated the FEM into the data term and solved it in an optimal control formulation. However, this approach necessitates manually identifying zero boundary conditions (ZBCs) and force locations.

Instead of incorporating the FEM as a data term or a strain/deformation term, Suwelack et al. \cite{suwelack2014physics} formulated the movement of the preoperative model to the intraoperative surface as an electrostatic–elastic problem, which the FEM solves. The preoperative model is assumed to be electrically charged and is attracted to the oppositely charged intraoperative surface by its electric potential field.



\textcolor{black}{Despite being in the era of deep learning, learning-based methods for liver deformation registration are still constrained by several factors: the scarcity of large datasets for training and testing, issues with interpretability, and suboptimal accuracy. Additionally, effectively combining learning-based methods with biomechanical model-based approaches to address registration problems still remains a challenge.} In learning-based techniques, the biomechanical model is often employed to simulate various deformations for training the neural network. Pfeiffer et al. \cite{pfeiffer2020non} proposed V2Snet (volume-to-surface registration network), designed to estimate the deformation of a volume mesh to an intraoperative surface. Tagliabue et al. \cite{tagliabue2021data} introduced BA-Net (binary-attachment network) to predict the locations of the attachment points during tissue dissection. The predicted attachment points can be used as ZBCs to update a patient-specific biomechanical model intraoperatively \cite{mendizabal2023intraoperative}. 

Although utilizing the FEM as strain energy might restrict deformation, it fails to ensure coherent movement of volumetric vertices to maintain the original geometry. On the other hand, the direct solution of the FEM-based data term requires the identification of Zero-Boundary Conditions (ZBCs) and force locations, as demonstrated in the existing method by Mestdagh et al. \cite{mestdagh2022optimal}.


Despite the numerous 3D-3D non-rigid registration methods that have been proposed, the liver registration community has limited access to their implementations, with only a few options available \cite{mestdagh2022optimal,khallaghi2015biomechanically,pfeiffer2020non}. Furthermore, there is a scarcity of publicly available liver registration datasets \cite{suwelack2014physics, brewer2019image, modrzejewski2019vivo}.





\subsection{Contributions}



\textcolor{black}{In this paper, we introduce a 3D-3D non-rigid registration method based upon incorporating a modified FEM into the surface matching/data term. Most non-rigid registration methods incorporate the FEM model into the strain energy/regularization term. Only a few methods [16, 19] include the FEM in the surface matching data term, which, to our best knowledge, requires the manual identification and specification of boundary conditions.}

\textcolor{black}{Here, we alleviate the need to prescribe boundary conditions by using a modified stiffness matrix of the FEM, which is incorporated into the data term. We use a well-known numerical method (diagonal loading) to stabilize the stiffness matrix. As a result, the modified data term enables us to not need to manually identify and prescribe boundary conditions, and also to not need to use an additional strain energy term to be minimized.}

To solve our formulation, we employ the Nesterov accelerated gradient algorithm and derive the necessary gradients. Additionally, we propose a novel optimal step size strategy that eliminates the need for manual step size tuning.

We validate our method using our both simulated and experimentally collected phantom datasets, as well as two publicly available datasets.  Comparative assessments are conducted against two open-source methods - a learning-based approach and an optimization-based method utilizing FEM for regularization, along with state-of-the-art closed-source methods. This straightforward yet effective approach consistently demonstrates its effectiveness relative to the benchmarks.

Given the limited availability of public datasets and accessible non-rigid registration methods, we aim to contribute to the field by publicly disseminating our constructed liver dataset and the developed registration algorithm, thereby advancing liver registration benchmarks.

\section{Preliminary Information}

Before liver surgery, patient-specific biomechanical finite element models are created using preoperative CT and/or MRI scans. These models are based on the patient's liver geometry derived from imaging scans, along with the material properties assigned to the tissue.  

The preoperative liver geometry is represented by a volumetric tetrahedral mesh model $\mathbf{\Omega}$ comprising $n$ nodes $\mathbf{x} =\{ \mathbf{x}_1,...,\mathbf{x}_n \vert \mathbf{x}_i \in \mathbb{R}^3\} \in \mathbb{R}^{3n}$. The preoperative liver surface is represented by a triangular mesh model $\mathbf{\partial\Omega}$ composed of $n_s$ nodes $\mathbf{x}_s \in \mathbb{R}^{3n_s}$, which is a subset of the volumetric mesh nodes. 

During surgery, the observed intraoperative surface is represented by a point cloud $\mathbf{y}=\{ \mathbf{y}_1,...,\mathbf{y}_m\vert \mathbf{y}_i \in \mathbb{R}^3\} \in \mathbb{R}^{3m}$. The intraoperative surface can be obtained from various sources, such as CBCT, an RGBD depth camera, structure light, stereo laparoscope, or manual swabbing with an optically tracked probe. The visible intraoperative surface varies in surgery settings \cite{clements2011organ, heiselman2018characterization}: for open surgery, the visible surface could reach over 50$\%$, while for laparoscopic surgery, typically, the extent of visible surface is much more limited, varying from 20 to 30$\%$.

The objective is to register a preoperative model with the deformed intraoperative surface data to map the sub-surface anatomical structures identified in the preoperative images into the intraoperative scene. A preliminary alignment of the preoperative model with the intraoperative surface is necessary before the non-rigid step, typically achieved through a rigid transformation. Existing IGLS systems commonly perform this alignment manually \cite{acidi2023augmented, golse2021augmented}; nevertheless, the manual alignment can be further improved with 
a variant of the ICP \cite{besl1992method, clements2008robust}.

\subsection{Biomechanical Model}
\label{sec:Biomechinical_model}

When surface forces $\mathbf{f}$ are applied to the liver boundary, the resulting displacements $\mathbf{u} =\{ \mathbf{u}_1,...,\mathbf{u}_n \vert \mathbf{u}_i \in \mathbb{R}^3\} \in \mathbb{R}^{3n}$  are a unique solution of the static equilibrium equation:

\begin{equation}\label{eqn:P(u)=f} 
\mathbf{P}(\mathbf{u}) = \mathbf{f}, 
\end{equation}

\noindent \textcolor{black}{where $\mathbf{P(u)} = \{P_1(\mathbf{u}),..., P_n(\mathbf{u})\}$ is a vector of nonlinear functions of $\mathbf{u}$ and are determined by the FEM model of the liver geometry and material properties. For a liver modeled by a linear elastic material, the forces are now a linear function of $\mathbf{u}$ where $\mathbf{P}(\mathbf{u}) =\mathbf{K}\mathbf{u}$ with $\mathbf{K}$ is a constant $3n$ x $3n$ stiffness matrix. For a nonlinear hyperelastic material, $\mathbf{P}(\mathbf{u})$ is a nonlinear function of the displacements \textbf{u}. The tangent stiffness matrix $\mathbf{K}_T$ defined as the derivative of $\mathbf{P}(\mathbf{u})$ with respect $\mathbf{u}$\
\begin{equation}\label{eqn:dp(u)_du} 
\frac{\partial \mathbf{f}}{\partial \mathbf{u}} = \frac{\partial \mathbf{P}(\mathbf{u})}{\partial \mathbf{u}} = \mathbf{K}_T(\mathbf{u}),
\end{equation}
is a matrix used in each iteration of the solution of a nonlinear problem, and it changes with each iteration. For a linear elastic material, the tangent stiffness matrix is simply the stiffness matrix.}
A more detailed description of the construction of the FEM models can be in the following book

In the forward problem, the displacement vector $\mathbf{u}$ throughout the domain $\mathbf{\Omega}$ can be determined if displacements and/or forces on the boundary $\mathbf{\partial\Omega}$ are known. The inverse registration problem solves for the boundary forces that generate a displacement field $\mathbf{u}^*$ that approximates the displacement field $\mathbf{u}$ between the partially observed intraoperative surface and preoperative model.

\subsection{Stiffness Matrix and Boundary Conditions}
\label{sec:Stiffness_Matrix_and_Boundary_Conditions}

The liver geometry and material properties are encoded in the stiffness matrix of the finite element model. This matrix characterizes the object's resistance to nodal deformation when subjected to nodal forces. In the case of a linearly elastic biomechanical model\cite{bonet1997nonlinear}, Young's modulus ($E$), and Poisson's ratio ($\nu$) account for the material's elastic properties.

The stiffness matrix, as formulated, is singular and cannot be inverted. This singularity indicates a scenario where the structure lacks defined boundary conditions or constraints, allowing the structure to undergo rigid motion without internal deformations. In such a state, applied forces can result in indeterminate displacements.

The imposition of boundary conditions, specifying the solution's values on the boundary, constrain rigid-body motions. Consequently, the stiffness matrix becomes invertible, enabling the determination of a unique solution for displacements resulting from applied forces. In the context of assessing liver deformation during surgery, it is crucial to identify the locations featuring ZBC, which typically correspond to the attachment points of the liver. This identification is vital for a meaningful and accurate organ behavior simulation.



\textcolor{black}{To construct a FEM in linear elastic version as an example, where \(\mathbf{f} = \mathbf{K}\mathbf{u}\), the values of \(E\) and \(\nu\) are used to construct the constant 3n x 3n stiffness matrix \(\mathbf{K}\) \cite{bonet1997nonlinear,kim2014introduction}. Given forces and ZBC, displacement \(\mathbf{u}\) can be solved.}

\section{Methods}\label{sec3}

\begin{figure*}[h]
\centering
\includegraphics[width=0.99\textwidth]{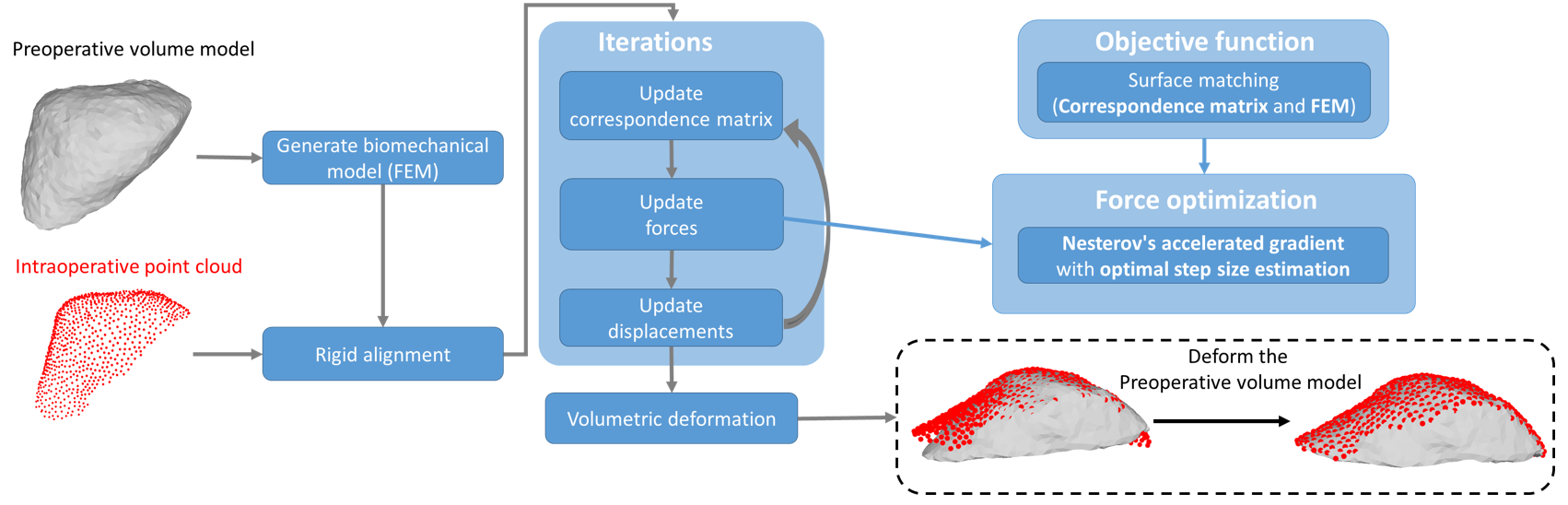}
\caption{ Overview of the proposed method. The process takes a biomechanical model generated from the volumetric liver model (typically segmented from a CT or MRI image) and an intraoperative point cloud. Force optimization leads to volumetric deformation, aligning the preoperative volume model with the intraoperative point cloud. }
\label{fig:Illustration}
\end{figure*}

Our method employs a force-driven approach to deform the source model for optimal alignment with the target point cloud. Fig. \ref{fig:Illustration} provides an overview of our registration method. The objective function comprises a surface matching term. The surface matching term utilizes a correspondence matrix generated by a closest point operator and a FEM. Minimization of the objective function is achieved through an accelerated gradient algorithm. During the force optimization phase, we initially determine the optimal step for calculating Nesterov's accelerated gradient \cite{Nesterov1983AMF}. Subsequently, the gradient is employed to update forces. In each iteration, we first update the correspondence matrix via the closest point operator, then update the force through force optimization; finally, we update the displacements using the FEM. \textcolor{black}{The FEM ensures that the displacement or deformation preserves the geometry of the preoperative model, maintaining the integrity of the node-to-node connections.} After a set number of iterations, the predicted volumetric deformation enables the preoperative volume model to align with the intraoperative point cloud. In the following section, we will detail each of these steps.

\subsection{Objective Function}
\label{sec:Ojbective_function}
We perform the registration by determining a surface force vector $\mathbf{f}$ that results in a displacement vector $\mathbf{u_f}$, which minimizes the objective function:


\begin{equation}\label{eqn:Objective function} 
\min_{\mathbf{f}\in\mathbf{F}}\Phi(\mathbf{f}) = J(\mathbf{u_f}),
\end{equation}



\noindent
where $J$ is a data term that measures the discrepancy between the predicted deformed preoperative surface $\mathbf{\partial\Omega}_\mathbf{u}$  and the observed intraoperative point cloud data. \st{}




The data term is given by
\begin{equation}\label{eqn:Data Term} 
J(\mathbf{u}) = \frac{1}{2}\|\mathbf{C{\color{black}(u)}}(\mathbf{x}+\mathbf{u})-\mathbf{y}\|^2,
\end{equation}
\noindent
where $\mathbf{C} \in  \mathbb{R}^{3m\times3n}$ is a soft correspondence matrix, $\mathbf{x} \in\mathbb{R}^{3n}$ are spatial location of the preoperative mesh nodes,   $\mathbf{u} \in\mathbb{R}^{3n}$ are displacements associated with the nodes, and $\mathbf{y} \in\mathbb{R}^{3m}$ is the intraoperative point cloud. The correspondence matrix $\mathbf{C}$ maps the deformed model to points on the intraop surface will be detailed in Sec. \ref{sec:Correspondence_matrix}. 

\textcolor{black}{The liver biomechanical model can be incorporated into Eq. \ref{eqn:Data Term} by using the relationship between $\mathbf{u}$ and $\mathbf{f}$ given by Eq.\ref{eqn:dp(u)_du}:}

\textcolor{black}{\begin{equation}\label{eqn:Data TermF} 
J(\mathbf{f}) = \frac{1}{2}\|\mathbf{C(u)}(\mathbf{x}+\mathbf{K}^{-1}\mathbf{f})-\mathbf{y}\|^2.
\end{equation}}\vspace{1mm}

\noindent \textcolor{black}{The resulting data term now becomes a function of the forces applied to the surface of the liver.}
Our data term is a reformulation of the data term from Mestdagh et al. \cite{mestdagh2022optimal, Mestdagh2022} where we have explicitly incorporated the correspondence matrix and the dependence of $J(\mathbf{u})$ on $\mathbf{u}$.

\subsection{Correspondence Matrix}
\label{sec:Correspondence_matrix}

The correspondence matrix $\mathbf{C}$ contains the correspondences between the points in the intraoperative point cloud and preoperative model. Ideally, if corresponding fiducial points or anatomical landmarks could be consistently identified in the pre- and intraoperative spaces, the correspondence matrix would be a one-to-one binary matrix. However, obtaining an accurate binary correspondence matrix is non-trivial unless manual annotation is used, which is highly user-dependent and prone to error.

Instead of a binary correspondence matrix, a soft correspondence matrix is often constructed, indicating the probabilistic correspondence between a point in one space and another. The elements $C_{ij}$ of the correspondence matrix $\mathbf{C}$ denote the probability that the intraoperative point $\mathbf{y}_i$ corresponds to the preoperative mesh point $\mathbf{x}_j + \mathbf{u}_j$. Each row of the correspondence matrix should sum to one.   

Surface matching methods such as the closest point operator and Gaussian mixture models \cite{khallaghi2015biomechanically, myronenko2010point} use the Euclidian distance between points in the pre-and intra-operative space as an alignment metric. Feature-based learning \cite{yang2023learning} or non-learning \cite{robu2018global} methods are used to measure the similarity between features in both spaces and can be used to construct a soft correspondence matrix.

Here, we use the closest point operator to construct a soft correspondence matrix. For a given intraoperative point $\mathbf{y}_i$, the closest point operator determines its closest point $\tilde{\mathbf{y}}_i$: 
\begin{equation}\label{eqn:Barycentric} 
\tilde{\mathbf{y}}_i = \lambda_{i}(\mathbf{x}_{i}+\mathbf{u}_{i}) +\lambda_{j}(\mathbf{x}_{j}+\mathbf{u}_{j}) + \lambda_{k}(\mathbf{x}_{k}+\mathbf{u}_{k}),
\end{equation} \vspace{1mm}
\noindent
on the preoperative surface, which is defined by the three nodes  $\mathbf{x}_{i,j,k}$, displacements vectors $\mathbf{u}_{i,j,k}$, and barycentric coordinates $\lambda_{i,j,k}$ associated with the surface triangle that contains the closest point.

The $\lambda_{i,j,k}$ sums to one, representing confidence values $C_{i,(i,j,k)}$. Thus, the 3m x 3n correspondence matrix is extremely sparse, where each row contains only three non-zero entries corresponding to the barycentric coordinates. The closest point operator and the resulting correspondence matrix are a function of the undeformed mesh $\mathbf{x}$, intraoperative point cloud $\mathbf{y}$, and the nodal displacements $\mathbf{u}$.




\subsection{Optimization}
\label{sec:optimzation}


Nesterov's accelerated gradient \cite{Nesterov1983AMF} is used to minimize the objective function Eq. \ref{eqn:Objective function}. \textcolor{black}{Nesterov’s accelerated gradient is a momentum-based optimizer that adds a fraction of the previous update to the current one, creating a momentum effect that accelerates the algorithm toward the minimum.} The following steps

\begin{equation}\label{eqn:proximalEq1}
\mathbf{p}^k = \mathbf{f}^k+\frac{k}{k+3}\Bigl(\mathbf{f}^k - \mathbf{f}^{k-1}\Bigr),
\end{equation}
\begin{equation}\label{eqn:proximalEq2} 
\mathbf{f}^{k+1} = \mathbf{p}^{k}-\alpha^k  \nabla J(\mathbf{p}^k),
\end{equation}

\noindent
are iterated until a stopping condition is met. $k$ represents the number of current iterations.


\textbf{Nesterov's accelerated gradient} is given by Eq. \ref{eqn:proximalEq1} and Eq. \ref{eqn:proximalEq2}. $\mathbf{p}^k$ carries momentum from the previous iteration {\color{black}by combining the force estimates of the current and previous iterations}. Eq. \ref{eqn:proximalEq2} updates the forces, where $\nabla J(\mathbf{p}^k)$ is the gradient of the data term  (Eq. \ref{eqn:Data TermF}) with respect to  $\mathbf{p}$, and $\alpha^k$ is the step size. $\nabla J(\mathbf{p}^k)$ is given by
\begin{equation}\label{eqn:gradient J0} 
{\color{black}\nabla J(\mathbf{p}^k)  =  \mathbf{K}_T^{-1}\Bigl((\mathbf{x}+\mathbf{u}^{k})^T\frac{\partial \mathbf{C}^T}{\partial \mathbf{u}}  + \\ 
\mathbf{C}^T  (\mathbf{C}(\mathbf{x} + \mathbf{u}^{k}-\mathbf{y}) \Bigr)},
\end{equation}


\noindent {\color{black}where for a linear elastic material $\mathbf{K}_T$ is a constant and for a nonlinear material $\mathbf{K}_T$ is a function of $\mathbf{u}^k$ and can be updated as needed.} 
The correspondence matrix is recalculated during the iterative registration using the current estimate of the displacements.
{\color{black}To simplify the calculation of $\nabla J(\mathbf{p}^k)$ we employ an iteration-lagging method which updates the correspondence matrix for each iteration using nodal displacements $\mathbf{u}^k$ computed from the previous iteration. The correspondence matrix is now considered fixed and its variation with respect to $\mathbf{u}$ can be neglected (i.e. $\frac{\partial \mathbf{C}^T}{\partial\mathbf{u}} = 0$). Now $\nabla J(\mathbf{p}^k)$ is given by
\begin{equation}\label{eqn:gradient J} 
\nabla J(\mathbf{p}^k)  =  \mathbf{K}_T^{-1}\mathbf{C}^T\Bigl(\mathbf{C}(\mathbf{x} + \mathbf{u}^k)-\mathbf{y} \Bigr).
\end{equation} }

\textbf{Optimal step size estimation}. In Nesterov's accelerated gradient, the step is usually fixed. To further accelerate the optimization, we determine the optimal step size $\alpha^k$ by substituting the gradient descent term (Eq. \ref{eqn:proximalEq2}) into the data term (Eq. \ref{eqn:Data Term}) and minimizing it with respect to $\alpha$: 
\begin{equation}\label{eqn:AlpObjectiveFunction}
\alpha^k=\arg \min_ {\alpha} \frac{1}{2}\Bigl\|\mathbf{C}\Bigl(\mathbf{x} + \mathbf{K}^{-1}_T(\mathbf{p}^k-\alpha\nabla J(\mathbf{p}^k))\Bigr)-\mathbf{y}\Bigr\|^2,
\end{equation}

Setting the derivative of Eq. \ref{eqn:AlpObjectiveFunction} for 
$\alpha$ to zero, we can obtain the optimal step size:
\begin{equation}\label{eqn:AlpUpdate}
\alpha^k=\frac{\Bigl (\mathbf{C}\mathbf{K}^{-1}_T\nabla J(\mathbf{p}^k) \Bigr )^T \Bigl(\mathbf{C}(\mathbf{x} + \mathbf{K}^{-1}_T\mathbf{p}^k)-\mathbf{y} \Bigr)}{\Bigl (\mathbf{C}\mathbf{K}^{-1}_T\nabla J(\mathbf{p}^k) \Bigr)^T \Bigl(\mathbf{C}\mathbf{K}^{-1}_T\nabla J(\mathbf{p}^k) \Bigr)}.
\end{equation} 

\subsection{Modifications for Unknown Zero Boundary Conditions and Force Locations}
\label{sect:Modifications}




As mentioned in Section \ref{sec:Stiffness_Matrix_and_Boundary_Conditions}, without boundary conditions, the stiffness matrix $\mathbf{K}$ is ill-conditioned and cannot be inverted to yield a unique, stable solution. However, determining ZBCs and applied force locations is challenging during liver surgery. 


To address these unknowns, we modify the stiffness matrix to stabilize the model. Inspired by the concept of soft springs used by the commercial simulation software SOLIDWORKS\textsuperscript{\textregistered} \cite{solidworks2021softspring}, we stabilize the model by adding a soft spring to each node. This corresponds to adding a small stiffness term to the diagonal terms of the stiffness matrix: 

\begin{equation}\label{eqn:SoftSpring} 
\mathbf{K} \gets  \mathbf{K} + k_{ss}\mathbf{I},
\end{equation}

\noindent
where $k_{ss}$ is the stiffness of the soft spring and $\mathbf{I}$ is the identity matrix. This is analogous to the work presented by Peterlik \textit{et al.} \cite{peterlik2018fast}, where a damping term was added to the stiffness matrix, which was then integrated into a strain/deformation energy term to regularize a data term. In our work, the modified stiffness matrix (Eq. \ref{eqn:SoftSpring}) is integrated directly into the date term via the FEM. The added small stiffness increases the inertia of the liver model being deformed, preventing rigid motion
without internal deformations. This approach enables us to allow the forces to be applied at any spatial location on the liver surface.




As a consequence of these modifications, we no longer need to quantify the \textcolor{black}{actual location of ZBC's and }physical forces applied to the liver's surface; rather, we empower the optimization algorithm to identify both the spatial distribution and magnitude of applied forces that optimally deform the preoperative model to match the intraoperative point cloud. \textcolor{black}{This modified tangent stiffness matrix is incorporated into the gradient of the our data term described in Eq. \ref{eqn:gradient J0}.}
 
This formulation is a key contributor to our goal of developing an algorithm that allows us to register the preoperative and intraoperative without prior knowledge. As we shall show, these modifications yield accurate registration results.

\section{Experimental Set-up}



\textbf{Datasets}. The experiments are conducted on the following datasets:


\subsubsection{\textit{In silico} phantom} 
\begin{figure}[h]
\centering
\includegraphics[width=0.49\textwidth]{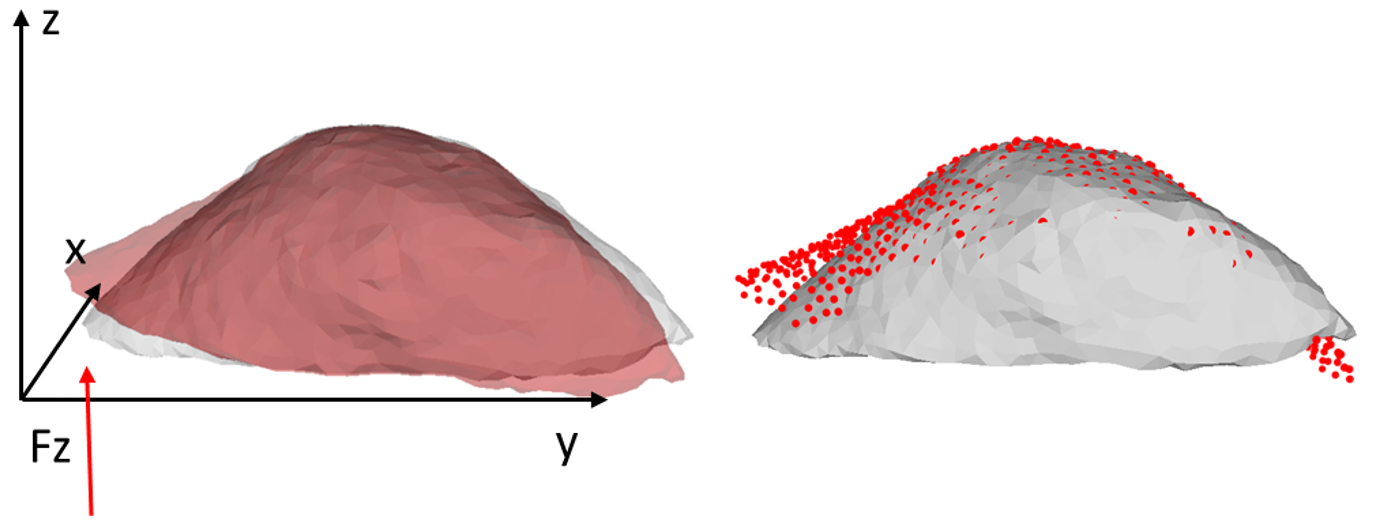}
\caption{The generation of an \textit{in silico} phantom. The \textit{in silico} phantom undergoes deformation with forces applied perpendicularly to the x-y plane, visualized by the deformed model in red. The cropped, deformed partial surface, represented as a red point cloud, constitutes the simulated intraoperative point cloud.}
\label{fig:simulated_phantom}
\end{figure}

As illustrated in Fig. \ref{fig:simulated_phantom}, we simulated the deformation of the \textit{in silico} phantom using a linear elastic model with Young's modulus $E=1$
and a Poisson's ratio $\nu=0.49$. Forces are applied perpendicular to the x-y plane on a liver mesh model obtained from OpenHELP \cite{kenngott2015openhelp}. The resulting deformed surface is cropped to generate the simulated intraoperative point cloud, consisting of 934 vertices. The volumetric preoperative mesh comprises 4291 vertices and 1919 tetrahedra.


\begin{figure}[h]
\centering
\includegraphics[width=0.49\textwidth]{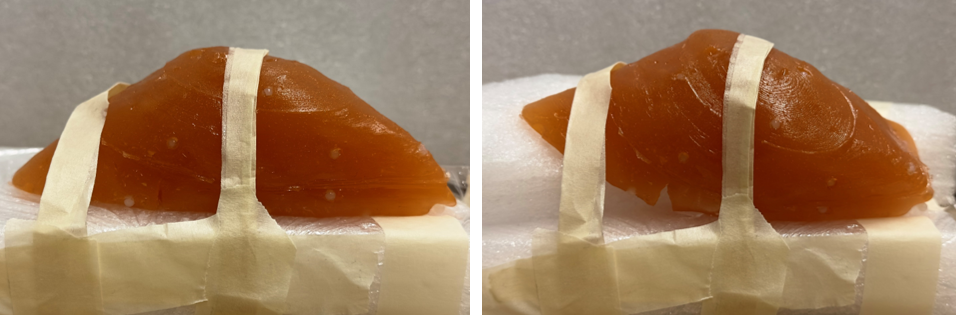}
\caption{ The generation of a  \textit{in vitro} silicone phantom. The undeformed liver phantom is shown on the left, and the deformed one on the right, with wedges added under its posterior surface.}
\label{fig:phantom}
\end{figure}

\begin{figure}[h]
\centering
\includegraphics[width=0.49\textwidth]{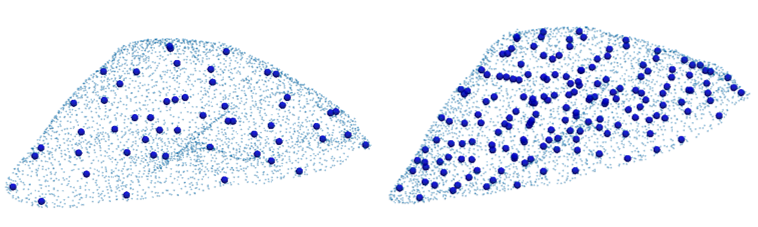}
\caption{Distributions of fiducial markers in \textit{in vitro} silicone phantoms A (left) and B (right). }
\label{fig:phantom_markers}
\end{figure}

\subsubsection{\textit{In vitro} phantoms} We created two liver phantoms, A and B, each embedded with 53 and 176 fiducial markers, respectively,  \textcolor{black}{as illustrated in Fig. \ref{fig:phantom_markers}}, and fabricated using \textcolor{black}{synthetic gelatin (Humimic Medical Gelatin \#0) and} a 3D-printed mold based on a liver model \cite{merrell2023developing}. To induce deformation, wedges featuring different gradients were strategically inserted underneath parts of the posterior side of the undeformed phantom, as depicted in Fig. \ref{fig:phantom} (b). The resulting phantom, with the wedge underneath, is referred to as the deformed phantom. We used the undeformed data as the preoperative data and the deformed data as the intraoperative data. The undeformed liver meshes, featuring around 9000 faces and 4000 vertices, were processed using TetGen \cite{hang2015tetgen} to generate the source volumetric meshes consisting of approximately 9000 volumetric vertices and 80000 tetrahedra.



Phantom A and Phantom B underwent two deformations, respectively, generating \textcolor{black}{four} deformed phantom configurations. We number the deformed phantoms (No. 1-2 from phantom A and No. 3-4 from phantom B). CT scans of both undeformed and deformed phantoms were acquired, and their surfaces and fiducial landmarks were manually segmented. Undeformed and deformed phantoms were manually aligned and refined with an ICP registration implemented using the Open3D Python library \cite{zhou2018open3d}. Initial registration errors are detailed in Table \ref{tab:kelly_dataset}. 

Additionally, partial anterior surfaces were cropped to achieve surface visibility ratios ranging from approximately $18\%$ to $27\%$, calculated as the ratios of cropped surface areas to the full surface areas, as summarized in Table \ref{tab:kelly_dataset}.


\subsubsection{Open-CAS dataset} The Open-CAS dataset contains three \textit{in silico} phantoms and one \textit{in vitro} phantom, as described in detailed in \cite{suwelack2014physics}. Deformations of the \textit{in silico} phantoms are generated using a non-linear biomechanical model. The \textit{in vitro} phantom featuring 6 fiducial markers was fabricated from silicone, and two intraoperative surfaces from a CT and stereo-endoscopic imaging are included.


\subsubsection{Sparse dataset} We use the Sparse Non-rigid Registration Challenge \cite{brewer2019image, collins2017improving,heiselman2023comparison}, where 
\textcolor{black}{112 samples with ground truth are released \cite{heiselman2024image}.}
Intraoperative surfaces are presented with sparse 3D point clouds collected using an optically tracked stylus. \textcolor{black}{The dataset also provides initial rigid transformations to align the preoperative and intraoperative datasets. Lastly, the dataset also includes sub-dataset splits featuring different visibility ratios and noise magnitudes, as described in detail in \cite{heiselman2023comparison}.}


\textbf{Evaluation}. The registration accuracy was estimated by calculating nodal displacement errors for the simulated cases and Target Registration Error (TRE) for the phantom cases. Their calculations share a common form:

\begin{equation}
\text{error}_i = \lVert \mathbf{Y}_i - \mathbf{W}_i(\mathbf{X}_i) \rVert_{2},
\end{equation}

\noindent
where $\mathbf{X}_i$ denotes a fiducial marker location or volumetric vertex position in the undeformed data, $\mathbf{Y}_i$ represents a fiducial marker or vertex position in the deformed data, $\mathbf{W}_i$ signifies the estimated transformation from a registration method for $\mathbf{X}_i$, and $\lVert \rVert_{2}$ denotes the Euclidean distance. Considering the four nearest points, the nearest neighbor interpolation is employed to propagate the estimated volumetric deformation to the fiducial markers.


\begin{figure*}[h]
\centering
\includegraphics[width=0.95\textwidth]{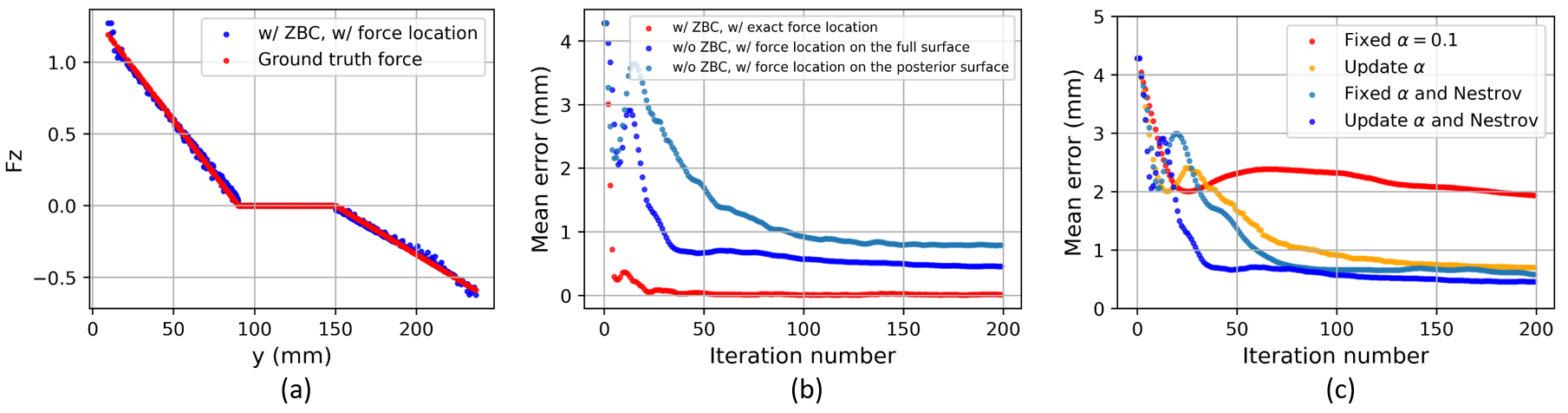}
\caption{Results of \textit{in silico} phantom study. Force recovery with correct boundary conditions (ZBCs) and force locations (left).
Examination of priors on registration results (middle). Optimization speed and accuracy through optimization step $\alpha$ update and Nesterov's acceleration (right).}
\label{fig:in_silico_phantom}
\end{figure*}

\begin{table*}[htpb]
\centering
\caption{Comparison of registration errors on the phantom dataset, presented as Mean ± Standard Deviation (Max) Error in millimeters. The lowest mean values and the lowest maximum values are highlighted in bold.}
\resizebox{0.99\linewidth}{!}{
\begin{tabular}{|c|c|c|c|c|c|c|}
\hline
Phantom No.  & Visibility ratio & Initial & Rigid Procrustes & V2S-net & GMM-FEM & Proposed Method \\
\hline
\multirow{ 2}{*}{1} 
& 26.22$\%$ 
& \multirow{ 2}{*}{8.31 $\pm$ 7.54 (32.76)}
& \multirow{ 2}{*}{3.31 $\pm$ 4.24 (20.47)}
&   5.07 $\pm$ 4.61 (21.41)
& 5.17 $\pm$ 3.95 (22.58)
& \textbf{1.62} $\pm$ 1.87 (\textbf{11.58})
\\

& 100$\%$ 
& 
& 
&    3.98 $\pm$ 3.87 (18.81) 
&  1.46 $\pm$ 2.06 (9.68)
& 
\textbf{1.29} $\pm$ 1.06 (\textbf{5.11})
\\
\hline
\multirow{ 2}{*}{2} 
& 20.79$\%$ 
& \multirow{ 2}{*}{11.50 $\pm$ 4.34 (22.43)}
& \multirow{ 2}{*}{5.52 $\pm$ 5.27 (23.92)}
&   9.03 $\pm$ 5.36 (25.23)
& 7.58 $\pm$ 3.63 (20.10)
& 
\textbf{1.55} $\pm$ 1.58 (\textbf{9.92})
\\

& 100$\%$ 
& 
& 
&    5.12 $\pm$ 3.44 (17.36) 
& 1.05 $\pm$ 1.28 (5.90)
&  \textbf{1.07} $\pm$ 0.83 (\textbf{4.17})



\\
\hline
 \multirow{ 2}{*}{3} 
& 21.24$\%$
& \multirow{ 2}{*}{5.36 $\pm$ 2.59 (19.21)}
& \multirow{ 2}{*}{2.34 $\pm$ 2.76 (17.82)}
&   3.97 $\pm$ 2.51 (18.75)
& 6.62 $\pm$ 3.48 (16.56)
& 
\textbf{1.03} $\pm$ 0.78 (\textbf{4.66})

\\

& 100$\%$ 
& 
& 
&    3.55 $\pm$ 2.26 (15.57) 
& 2.31 $\pm$ 2.05 (10.29) 
& 
\textbf{0.44} $\pm$ 0.21 (\textbf{1.20})

\\
\hline
 \multirow{ 2}{*}{4} 
& 19.36$\%$ 
& \multirow{ 2}{*}{10.73 $\pm$ 3.57 (19.16)}
& \multirow{ 2}{*}{1.32 $\pm$ 1.49 (9.78)}
&   5.95 $\pm$ 2.39 (16.86)
& 17.54 $\pm$ 3.68 (24.10)
&  \textbf{0.77} $\pm$ 0.72 (\textbf{5.55})

\\

& 100$\%$ 
& 
& 
&    2.31 $\pm$ 1.19 (9.27) 
&  1.15 $\pm$ 1.62 (8.14) 
& \textbf{0.35} $\pm$ 0.16 (\textbf{0.87})
\\
\hline
\end{tabular}
}
\label{tab:kelly_dataset}
\end{table*}

\begin{figure*}[h]
\centering
\includegraphics[width=0.99\textwidth]{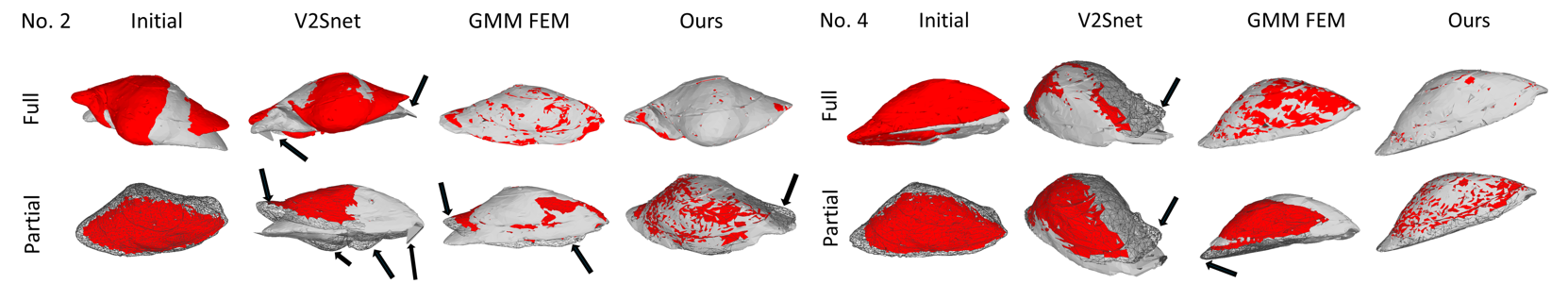}
\caption{ Qualitative comparison of the registration results yielded by different methods for Phantom datasets No. 2 and No. 4, presenting both initial undeformed and estimated deformed meshes yielded by all the methods. For cases with partial data, the edges of the complete deformed meshes are also displayed. The intraoperative point cloud is highlighted in red, and obvious disparities between the mesh and intraoperative point cloud are marked with arrows.}
\label{fig:Kelly_examples}
\end{figure*}

\textbf{Implementation}. The proposed method was implemented in C++ using the Eigen library \cite{eigenweb}, running on a 3.60 GHZ Intel i9-9900K CPU and 64 GB of RAM. \textcolor{black}{The Cholesky decomposition implemented in the Eigen library, named SimplicialLLT, is used to invert the stiffness matrix from Eq. \ref{eqn:SoftSpring}. An in-house implementation of a finite element model was used for the biomechanical calculations.} We use a linear elastic biomechanical liver model for all experiments. Since a linear biomechanical model is employed, Young's modulus $E$ solely influences the scale of the estimated forces \textcolor{black}{and does not affect the displacement solution}. For this study, we scale the force by setting $E$ to 1. Poisson's coefficient is set as $\nu = 0.49$, a typical value for modeling incompressible tissue \cite{sparks2008constitutive}. The soft spring constant was set to $k_{ss}=0.01$, and the optimization algorithm was stopped at 200 iterations. \textcolor{black}{The implemented procedure is detailed in Algorithm 1.}



\textbf{Open source methods for comparison}. We use the official implementations of V2Snet\cite{pfeiffer2020non} and GMM-FEM \cite{khallaghi2015biomechanically} as benchmarks for comparison. For V2Snet, we use the official weight included in the implementation. GMM-FEM \cite{khallaghi2015biomechanically} is an optimization-based method that incorporates the FEM as a strain energy term to regularize the CPD \cite{myronenko2010point}. We implemented it using its suggested parameter values, which were tested in our \textit{in silico} phantom described in Section \ref{sec:in_silico} and gave the lowest errors. Furthermore, we include the Rigid Procrustes method that uses RANSAC ICP implemented in the Open3D library \cite{zhou2018open3d} with ground truth correspondences to estimate the lowest errors that may be achieved via the rigid registration.

\algnewcommand{\Inputs}[1]{%
  \Statex \textbf{Inputs:}
  \Statex \hspace*{\algorithmicindent}\parbox[t]{.9\linewidth}{\raggedright #1}
}

\algnewcommand{\Outputs}[1]{%
  \Statex \textbf{Outputs:}
  \Statex \hspace*{\algorithmicindent}\parbox[t]{.9\linewidth}{\raggedright #1}
}

\def\NoNumber#1{{\def\alglinenumber##1{}\State #1}\addtocounter{ALG@line}{-1}}

\begin{algorithm}
\caption{Linear Elastic Surface Matching Registration}
\begin{algorithmic}[1]
\Inputs{ \textbf{x} the preoperative (source) nodes of liver mesh \newline \textbf{y} the intraoperative (target) surface point cloud 
\newline $k_{ss}$ the soft spring constant \newline maxIters the maximum number of iterations\vspace{1.0mm}
}
\vspace{0.5mm}
\Outputs{\textbf{u} displacements for registered liver mesh}
\setcounter{ALG@line}{0}

\State initialize displacements $\mathbf{u^0} \gets 0$;
\State initialize forces $\mathbf{f^0} \gets 0$;
\State calculate stiffness matrix $\mathbf{K} \gets \mathbf{K} + k_{ss}$;
\For{k = 0 to maxIters}
\State update mesh nodal displacements $\mathbf{u}^k =\mathbf{K}^{-1} 
\mathbf{f}^k$;
 \State update mesh nodes locations $\mathbf{x} \gets \mathbf{x} + \mathbf{u}^k$;
 \State compute correspondence matrix \textbf{C};
 \State update momentum term $\mathbf{p}^k$ (Eq.\ref{eqn:proximalEq1});
 \State compute gradient of data term $\nabla J(\mathbf{p}^k)$ (Eq.\ref{eqn:gradient J});
 \State compute optimal step size $\alpha^k${ (Eq.\ref{eqn:AlpUpdate})};
 \State update forces ${\mathbf{f}}^{k+1}$ (Eq.\ref{eqn:proximalEq2});
 \EndFor
\end{algorithmic}
\end{algorithm} 

\section{Results}

\subsection{\textit{In Silico} Phantom Validation}
\label{sec:in_silico}





We initiate our study using a noise-free and fully controlled \textit{in silico} liver phantom to understand how our algorithm performs under various settings. 

We first examine the algorithm's ability to predict ground truth forces when zero boundary conditions and force locations are known. 
When provided with precise material parameters, exact force locations, and zero boundary conditions (ZBC), our method effectively recovers forces that closely approximate the ground truth forces, as illustrated in Fig. \ref{fig:in_silico_phantom} (a). This accurate force recovery translates to minimal registration errors, lower than 2 mm, as depicted in Fig. \ref{fig:in_silico_phantom} (b). 

Next, we examine a more realistic situation where prior knowledge of the location of zero boundary conditions and forces is unknown. We eliminate ZBCs and include the soft spring constant $k_{ss}$. We explore two scenarios: the former assumes that the forces exist solely on the posterior surface, aligning with the assumption in \cite{rucker2013mechanics}; the latter assumes that forces can be distributed across the entire surface. As illustrated in Fig. \ref{fig:in_silico_phantom} (b), allowing the forces to be applied across the full liver surface yields superior registration results compared to confining the applied forces to the posterior surface.  The results also indicate accurate registration results can be obtained without prior knowledge of force locations and ZBCs. As noted before, we are no longer computing the actual physical forces applied to the liver's surface, but we rather empower the optimization algorithm to determine the fictitious forces that minimize the registration error. Subsequent experiments are conducted, assuming that forces can exist on the full surface.

Furthermore, we show that incorporating Nesterov's acceleration and updating $\alpha$ enhances the optimization process, resulting in faster convergence and improved accuracy, as depicted in Fig. \ref{fig:in_silico_phantom} (c). 

The computation depends on the number of FEM nodes and intraoperative point cloud, as our method relies on the FEM model and the closest point operator. \textcolor{black}{The overall computation time for this dataset is 22 seconds over the 200 iterations.}

\subsection{\textit{In vitro} Phantom Validation}
\label{sec:phantoms}


Table \ref{tab:kelly_dataset} summarizes the results for the phantom dataset, comparing the whole and partial surface registration errors of our proposed method relative to those achieved by V2S and GMM-FEM. Our method features a mean TRE below 5 mm across all testing samples. Notably, the differences in mean TREs between our predictions for partial and full surfaces are negligible, typically less than 1 mm. In contrast, GMM-FEM tends to exhibit lower errors on full surfaces, as illustrated by the mean and max TRE on the full surface of No. 4, which are 1.15 $\pm$ 1.62 (8.14) mm, compared to the partial surface, featuring an error of 17.54 $\pm$ 3.68 (24.10) mm. V2Snet encounters challenges in accurately predicting results for No. 2, resulting in a mean TRE exceeding 5 mm and a max TRE larger than 15 mm.


Fig. \ref{fig:Kelly_examples} illustrates several qualitative results that provide the reader with a deeper and more tangible interpretation of the registration performance achieved by our proposed and comparative methods. Notably, blade-like structures are evident in the deformed meshes generated by V2Snet, as indicated by arrows in both No. 2 and No. 4. Specifically, for both the full and partial surface cases of the No. 2 phantom, the estimated deformed mesh of V2Snet exhibits discrepancies in fully aligning with the ridge of the target surface. This misalignment is also noticeable in the estimation by GMM-FEM, particularly in the case of the partial surface for the No. 2 phantom. In contrast, our method consistently produces reasonable deformations for both the full and partial cases of the No. 2 phantom. 


\subsection{Open-CAS Dataset }
\label{sec:Open-CAS}

\begin{table*}[t!]
\centering
\caption{Comparison of registration errors on the Open-CAS dataset. Mean ± Standard Deviation (Max) Error in millimeters is presented. The lowest mean values and the lowest maximum values are highlighted in bold.}
\resizebox{0.99\linewidth}{!}{%
\begin{tabular}{|c|c|c|c|c|c|c|}
\hline
& Visibility & Initial  & Rigid Procrustes & V2S-net & GMM-FEM & Proposed Method \\
\hline
Simulation 1 
& 55.60$\%$
& 9.97 $\pm$ 8.28 (35.03)
&  5.88 $\pm$ 6.82 (25.36)
&   10.92$\pm$ 9.29 (38.31) 
& 1.90 $\pm$ 1.21 (\textbf{9.21}) 
& \textbf{1.61} $\pm$ 1.16  (9.82)
\\
Simulation 2 
& 58.68$\%$
& 7.47 $\pm$ 7.56 (29.22)
& 6.45 $\pm$ 26.13 (7.61)
&   3.06$\pm$ 3.25 (29.22)    
& 1.98 $\pm$ 1.36, (9.87) 
& \textbf{1.41} $\pm$ 1.14  (\textbf{5.99})

\\
Simulation 3 
& 63.46$\%$
& 9.17 $\pm$ 11.79 (56.68)
& 9.16 $\pm$ 13.62 (57.77)
&   4.91 $\pm$ 5.87 (56.68) 
&  1.80 $\pm$ 1.09 (\textbf{7.23}) 
&  \textbf{1.62} $\pm$ 1.23  (8.23)
\\
\hline
Phantom (partial CT) 
& 19.01$\%$
& \multirow{ 2}{*}{23.86 $\pm$ 15.07 (46.59)} 
& \multirow{ 2}{*}{13.95 $\pm$ 9.34 (25.95)} 
&   6.82 $\pm$ 3.40 (11.50)
&  6.80 $\pm$ 3.61 (12.40)
& \textbf{4.00} $\pm$ 1.37 (\textbf{6.23})
\\
Phantom (Stereo) 
& $\sim$  $20\%$
&  
& 
&    12.88 $\pm$ 2.90 (16.68)
&  15.16 $\pm$ 3.59 (21.56)
&  \textbf{11.64} $\pm$ 1.37  (\textbf{14.26})

\\
\hline

\end{tabular}
}
\label{tab:PBSM}
\end{table*}


The results are summarized in Table \ref{tab:PBSM}. In the simulation dataset, featuring over $50\%$ visibility, both GMM-FEM and our method exhibit comparable performance, with differences in mean TRE of less than 1 mm. Notably, for Simulation 2, our method achieves a maximum TRE of around 6 mm, significantly lower than the 10 mm achieved by GMM-FEM. V2Snet struggles to generalize on Simulation 1 and demonstrates a maximum TRE exceeding 20 mm for Simulations 2 and 3. The partial CT and stereo-reconstructed surfaces present approximately $20\%$ partial visibility, posing a challenge. In this scenario, our method outperforms all others, yielding the lowest mean and max TRE.

\begin{figure}[h]
\centering
\includegraphics[width=0.49\textwidth]{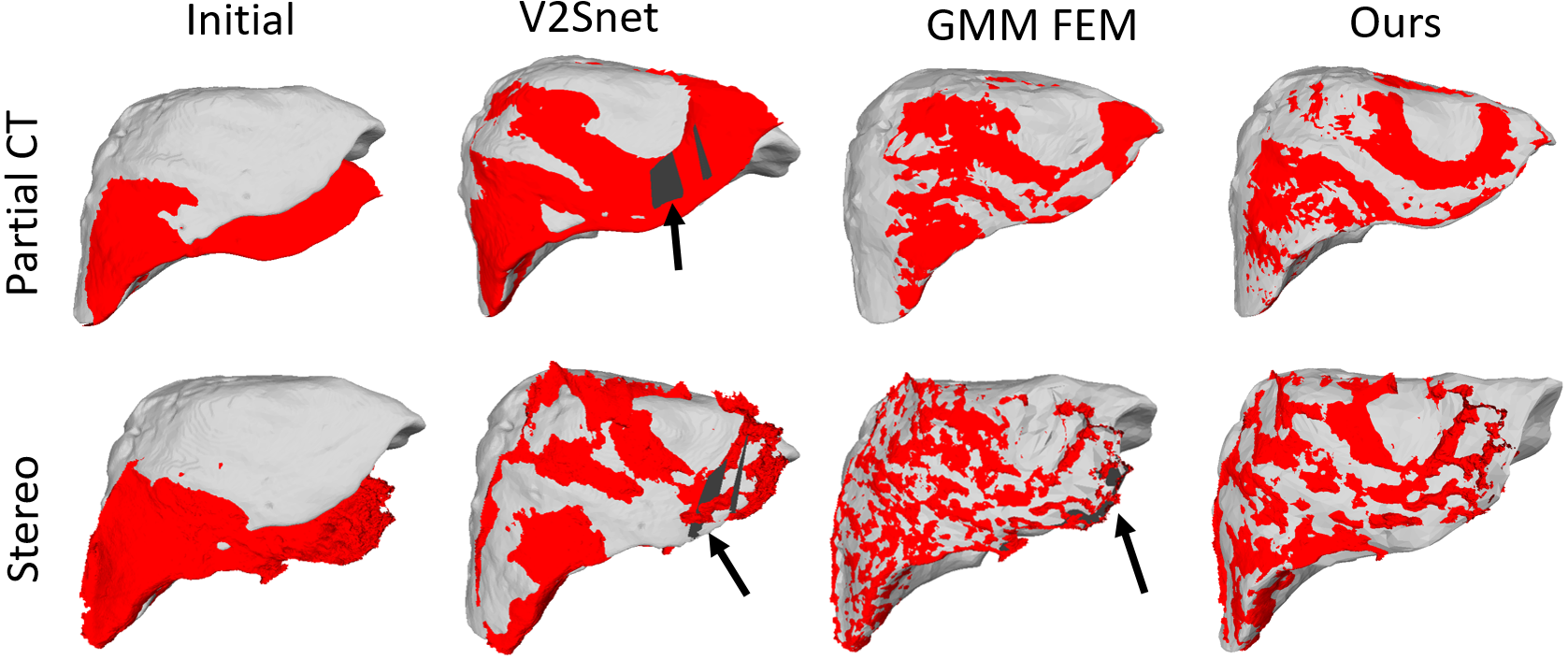}
\caption{Qualitative results achieved using the \textit{in silico} phantom cases of the Open-CAS dataset, showing both the initial and estimated meshes. The intraoperative point cloud is highlighted in red, and obvious disparities between the registered preoperative mesh and intraoperative point cloud are marked with arrows.}
\label{fig:PBSM}
\end{figure}

The qualitative results in Fig. \ref{fig:PBSM} support these findings. As observed previously, V2Snet generates deformed meshes featured with blade-like faces, evident in both cases in Fig. \ref{fig:PBSM}. GMM-FEM exhibits a similar phenomenon for the deformed mesh on the stereo-reconstructed surface. In contrast, our method consistently produces deformed meshes that maintain a realistic shape.

\subsection{Sparse Dataset}
\label{sec:sparse_data}

\begin{table*}[t!]
\centering
\caption{Comparison of registration errors on the Sparse dataset. Mean ± Standard Deviation (Median) of Mean Error in millimeters is presented. The lowest mean values and the lowest median values are highlighted in bold.}
\resizebox{0.99\linewidth}{!}{%

\begin{tabular}{|c|c|ccc|c|}
\hline
\multirow{2}{*}{Method}    & \multirow{2}{*}{Type}                 & \multicolumn{3}{c|}{Visibility}                  & \multirow{2}{*}{All} \\ 
                             &  & \multicolumn{1}{c}{20-28\%} & \multicolumn{1}{c}{28-36\%} & 36-44\%  &  \\ \hline
V2S-net \cite{pfeiffer2020non}                    
& \multirow{2}{*}{Deep learning}        
& \multicolumn{1}{c|}{7.79       $\pm$ 4.86      (6.61)} 
& \multicolumn{1}{c|}{ 6.79       $\pm$ 4.63      (6.10)} 
&   8.24       $\pm$ 6.34      (6.86)
& 7.55 $\pm$ 5.28 (6.37)                     \\  
Jia et al. \cite{jia2021improving}                       
&  
& \multicolumn{1}{c|}{ 4.22       $\pm$ 0.96      (4.02)}        
& \multicolumn{1}{c|}{4.57      $\pm$ 2.79      (3.88)}        
&             4.02       $\pm$1.31      (3.72)
& 4.29 $\pm$ 1.93 (3.84) \\ \hline
Heiselman et al. \cite{heiselman2020intraoperative}                
& 
& \multicolumn{1}{c|}{3.27       $\pm$ 1.07      (\textbf{2.89})} 
& \multicolumn{1}{c|}{3.00       $\pm$ 0.67      (2.97)} 
&    2.99 $\pm$ 0.78 (\textbf{2.73}) 
&  3.08 $\pm$ 0.85 (2.89)                    \\  
Mestdagh et al. \cite{mestdagh2022optimal}                     
&  
& \multicolumn{1}{c|}{3.54       $\pm$ 1.11      (3.47)}        
& \multicolumn{1}{c|}{ 3.27       $\pm$ 0.85      (3.19)}        
&  3.13       $\pm$ 0.82      (3.19)            
&  3.31 $\pm$ 1.86 (2.95)\\ 
Ringel et al.  \cite {ringel2023comparing}                    
&  Biomechanical
&   \multicolumn{1}{c|}{4.56 $\pm$ 0.88 (4.48)}  
&   \multicolumn{1}{c|}{4.62 $\pm$ 0.99 (4.71)}
&  4.62 $\pm$ 0.98 (4.37)      
&  4.60 $\pm$ 0.95 (4.52)	\\  
AIC \cite{heiselman2024optimal}                
&  model
&   \multicolumn{1}{c|}{5.06 $\pm$ 0.92 (4.93)}       
&   \multicolumn{1}{c|}{4.91 $\pm$ 1.11 (4.75)}    
&  4.61 $\pm$ 1.14 (4.52)       
&  4.86 $\pm$ 1.07 (4.77)	\\ 
GMM-FEM \cite{khallaghi2015biomechanically}                   
& (linear elasticity)
&  \multicolumn{1}{c|}{5.40       $\pm$ 1.56      (5.14) } 
& \multicolumn{1}{c|}{5.14       $\pm$ 1.22      (4.83)}        
& 4.55       $\pm$ 1.16      (4.58)                 
& 5.00       $\pm$ 1.37      (4.75) \\  
Proposed method                          
&  
& \multicolumn{1}{c|}{\textbf{3.05      $\pm$ 0.75} (2.95)}       
& \multicolumn{1}{c|}{\textbf{2.94      $\pm$ 0.66      (2.74)}}        
&   \textbf{2.78      $\pm$ 0.68}      (2.82)             
& \textbf{2.93      $\pm$ 0.68      (2.82)}
 \\ \hline
\end{tabular}

}
\label{tab:whole_sparse}
\end{table*}

\begin{table}[h]
\centering
\caption{Sensitivity analysis of the proposed method to the initial alignment analysis performed on the sparse dataset. Mean ± Standard Deviation (Median) of Mean Error in millimeters is presented. 
}
\resizebox{0.99\linewidth}{!}{%
\begin{tabular}{|c|c|c|}
\hline
  & Initial error & w/ proposed method \\
\hline
Optimal rigid registration 
& 3.77 $\pm$ 2.40 (3.16)
& 2.97      $\pm$ 0.66      (2.92)

    \\
Manually initialized ICP
& 5.53 $\pm$ 3.43 (4.84)
&  2.93     $\pm$  0.68      (2.82)
   \\
wICP \cite{clements2008robust}
&  7.73 $\pm$ 4.48 (6.74) 
& 2.97      $\pm$ 0.72      (2.85) 

    \\
\hline
\end{tabular}
}
\label{tab:initial_pose}
\end{table}

\begin{table}[h]
\centering
\caption{Performance of the proposed method to noise-free and noise-affected sparse datasets. Mean ± Standard Deviation of Mean Error in millimeters is presented.  }
\resizebox{0.95\linewidth}{!}{%
\begin{tabular}{|c|c|c|}
\hline
  & Noise = 0 mm & Noise = 2mm \\
\hline
Manually initialized ICP
& 5.53 $\pm$ 3.43 
&  5.59     $\pm$  1.61    \\
\hline
w/ Heiselman et al. \cite{heiselman2020intraoperative} 
& 2.50 $\pm$ 0.31 
&  3.34     $\pm$  0.87      \\
\hline
w/ Proposed method
& 2.23   $\pm$    0.34 
&  3.17   $\pm$    0.66  \\
\hline
\end{tabular}
}
\label{tab:noise}
\end{table}

\begin{figure*}[h]
\centering
\includegraphics[width=0.99\textwidth]{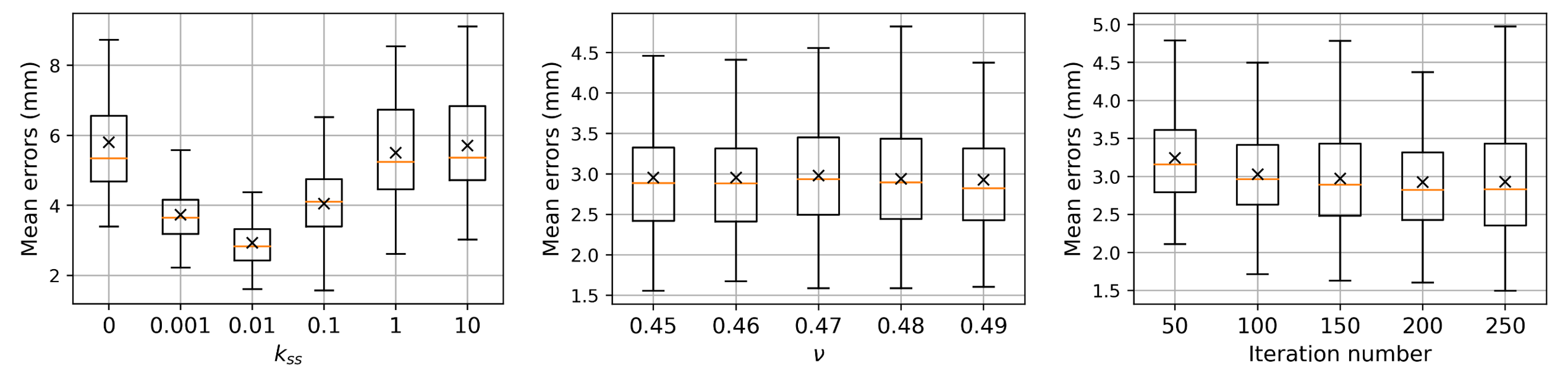}
\caption{Parameter sensitivity analysis for $k_{ss}$, $\nu$, and iteration number performed on the \textcolor{black}{entire Sparse dataset.} The median and mean values of \textcolor{black}{mean errors} are represented by orange lines and crosses, respectively. }
\label{fig:parameter_sparse}
\end{figure*}


\textcolor{black}{The registration errors are detailed in Tables \ref{tab:whole_sparse}, \ref{tab:initial_pose}, and \ref{tab:noise}. Except for GMM-FEM and our proposed method, results from other approaches align with previous reports \cite{heiselman2024optimal,heiselman2023comparison,heiselman2024image}. The other methods are either closed-source \cite{jia2021improving, heiselman2020intraoperative, ringel2023comparing}, require additional data processing \cite{pfeiffer2020non}, or involve manual input \cite{mestdagh2022optimal}. The most similar method to our proposed method is that of Mestdagh \textit{et al.} \cite{mestdagh2022optimal}, which also uses a FEM as the data term but requires manual boundary condition identification, which is known to pose significant challenges. Lastly, all biomechanical-based methods employ the linear elastic model.} 




\textcolor{black}{The sparse dataset includes three initial transformations for the rigid alignment. Table \ref{tab:whole_sparse} and Table \ref{tab:noise} report results using the manually initialized ICP transformation, consistent with the performance shown by Heiselsman \textit{et al.} \cite{heiselman2020intraoperative}. Table \ref{tab:initial_pose} presents results from our proposed method using all initial transformations, demonstrating its robustness across different starting conditions. The proposed method consistently achieves the lowest mean registration errors across various settings, including the entire sparse dataset (Table \ref{tab:whole_sparse}), different visibility conditions (Table \ref{tab:whole_sparse}), multiple initial transformations (Table \ref{tab:initial_pose}), and both noise-free and noise-affected scenarios (Table \ref{tab:noise}), demonstrating its effectiveness and robustness. As shown in Table III, our proposed method outperforms the method developed by Mestdagh \textit{et al.} \cite{mestdagh2022optimal}, highlighting its effectiveness in automatically identifying boundary conditions. Compared to other FEM-based methods \cite{heiselman2020intraoperative, khallaghi2015biomechanically}, our approach also achieves lower registration errors (Table III and Table V). Lastly, the learning-based methods \cite{jia2021improving, pfeiffer2020non} show less satisfactory performance than the best biomechanical approaches.}

\begin{figure}[h]
\centering
\includegraphics[width=0.49\textwidth]{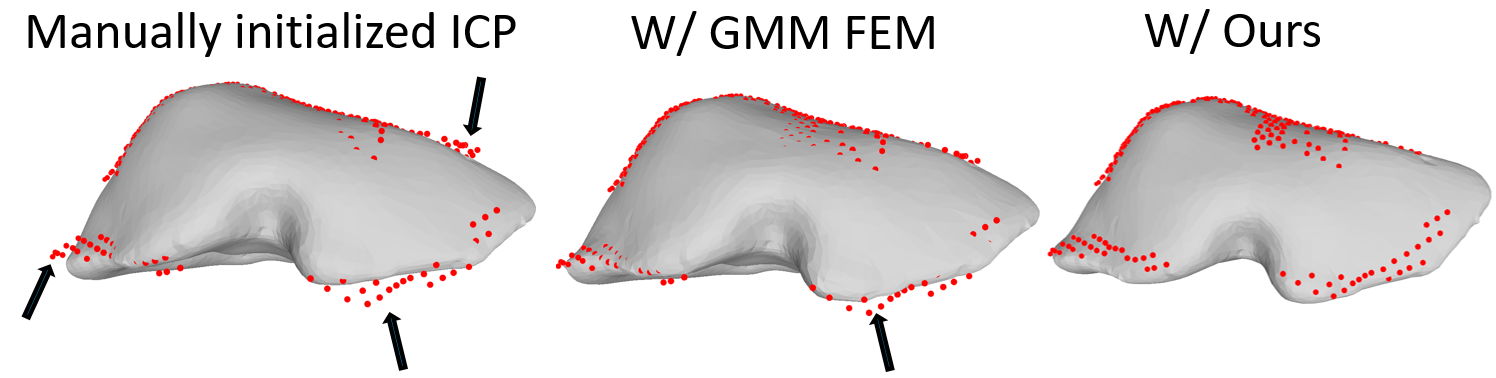}
\caption{Qualitative results for \textcolor{black}{Set0100 of the Sparse dataset, presenting the initial position from the manually initialized ICP, the official transformation,} and estimated deformed meshes. The intraoperative point cloud is highlighted in red, and obvious disparities between the mesh and intraoperative point cloud are marked with arrows.}
\label{fig:Sparse}
\end{figure}

Qualitative results for \textcolor{black}{Set100} in Fig. \ref{fig:Sparse} reveal challenges for GMM-FEM in aligning the right inferior ridges of the liver, where our method excels in accurately fitting the intraoperative point cloud.


\subsection{Parameter Sensitivity Analysis}
\label{sec:parameter_sensitivity}




The proposed method relies on three parameters: (1) $k_{ss}$, a stabilizing parameter for the solution; (2) Poisson's ratio $\nu$, defining the compressibility of the tissue; \textcolor{black}{and (3) iteration number.} To understand the sensitivity of our method to these parameters, we vary each parameter individually while keeping the others constant.  We examined their impact on registration error and conducted statistical Kolmogorov–Smirnov tests at a significance level of $\alpha= 0.05$,  \textcolor{black}{using the entire sparse dataset}. The results are shown in Fig. \ref{fig:parameter_sparse}.

\textcolor{black}{The registration exhibit a degree of sensitivity to the parameter $k_{ss}$. The default setting of $k_{ss}=0.01$ outperforms other settings ($p < 0.001$). The proposed method is not sensitive to Poisson's ratio $\nu$; hence, no statistical significance is found.} The default setting of Poisson's ratio $\nu=0.49$ corresponds to nearly incompressible tissue behavior. While some prior studies \cite{rucker2013mechanics, heiselman2020intraoperative} adopt $\nu=0.45$, our experiments involving variations from 0.45 to 0.49 show stable performance. Lastly, when we reduce the iteration number to 50, statistical significance is found ($p < 0.001$), suggesting the default iteration number 200 could be further tuned to reduce computation.


\section{Discussion}

\subsection{Summary of Main Findings}

We first introduce a soft spring modification to the stiffness matrix to eliminate the need for priors. This modification has been demonstrated to substantially reduce registration errors, as evidenced in Fig \ref{fig:parameter_sparse}. Secondly, the method is designed to search for optimal forces autonomously. In the absence of manually identified ZBC and force locations, this strategy is more beneficial than the previous approaches \cite{rucker2013mechanics} of constraining force locations to predefined areas, such as the posterior surface, shown in Fig. \ref{fig:in_silico_phantom} (b). In addition, our proposed optimal step determination, as demonstrated in Fig. \ref{fig:in_silico_phantom} (c), eliminates the need for manual tuning of the step size parameter and leads to quicker convergence and lower error.




We have evaluated our method across various datasets, encompassing our phantom datasets, the Open-CAS dataset, and the Sparse dataset. Moreover, in comparisons with two alternative \textcolor{black}{open source} methods (V2Snet and GMM-FEM) and \textcolor{black}{ closed source methods reported in the Sparse Data Non-rigid Registration Challenge \cite{heiselman2023comparison, brewer2019image}}, our approach consistently outperforms or shows comparable performance, as elaborated in Sections \ref{sec:phantoms}, \ref{sec:Open-CAS}, and \ref{sec:sparse_data}.

\textcolor{black}{In comparison with open source methods,} while GMM-FEM incorporates a FEM into the strain energy, regularizing the deformation field estimated by the CPD registration method, it does not guarantee faithful preservation of the original geometry (i.e., vertex connectivity) of the deformed surface. This becomes apparent in situations with noisy intraoperative point cloud data, as illustrated in Fig. \ref{fig:PBSM}. In contrast, our approach, where FEM inherently guides the deformation, ensures the maintenance of volumetric vertex connectivity, preserving the geometry during deformation. Furthermore, as demonstrated in Section \ref{sec:phantoms}, GMM-FEM exhibits sensitivity to the visibility of the intraoperative surface, as highlighted in its original paper \cite{khallaghi2015biomechanically}. V2Snet utilizes FEM to construct a training dataset \cite{pfeiffer2020non}. While it shows competitive results in specific instances, its overall performance and generalization capabilities require improvement. Nevertheless, the use of learning-based methods, offering prior information, has the potential to assist optimization methods in achieving faster and more accurate results.

\textcolor{black}{The sparse dataset enables comparison of our proposed method with other closed-source methods, particularly regarding surface visibility, initial alignment, and noise magnitude. Our method demonstrates robustness to these factors within reasonable variations and consistently achieves the lowest registration errors, as shown in Table \ref{tab:whole_sparse}, \ref{tab:initial_pose}, and \ref{tab:noise}.
}


\textcolor{black}{The proposed method is straightforward and effective, in contrast to registration methods driven by both surface-matching and strain energy terms. Our approach relies solely on the surface-matching term embedded within the FEM, eliminating the need to iteratively determine boundary conditions \cite{heiselman2020intraoperative, rucker2013mechanics} or manually specify them \cite{mestdagh2022optimal}. Additionally, our method depends on only a few parameters, making it easy to adapt to other datasets. Section \ref{sec:parameter_sensitivity} provides a sensitivity analysis of these parameters.  The stabilizing parameter $k_{ss}$ should be set carefully, though we recommend against extensive tuning, as the default parameter value performs well across the samples included in this study. Potential future work includes automating the process of determining the $k_{ss}$ value.}

\subsection{Linear Elastic Model and Deformation}

\textcolor{black}{ We have shown that our algorithm's linear elastic model version is parameter robust and leads to accurate registrations for datasets with deformations typically observed in laparoscopic \cite{heiselman2018characterization} and open liver surgery \cite{clements2011organ}. The linear elastic model is generally employed by most state-of-the-art methods \cite{mestdagh2022optimal, heiselman2020intraoperative, heiselman2023comparison}. In fact, all biomechanical-based methods reported in the Sparse Challenge dataset adopted the linear elastic model. The maximum TRE reported in the sparse data is 11.9 $\pm$ 3.7, consistent with clinically observed deformation magnitudes for open liver surgery reported in
Ref. \cite{clements2011organ, heiselman2024image}. }

\textcolor{black}{Liver deformation in laparoscopic interventions can be larger due to the creation of pneumoperitoneum. The Open-CAS dataset features even more significant deformations than the Sparse Challenge dataset. Even following the initial rigid alignment correction, mean TRE still ranges from 5.88 mm to 13.95 mm, which reflects the typical deformations observed in laparoscopic liver surgeries reported in \cite{heiselman2018characterization}, where surface deformations ranged from 7.7 mm to 12.6 mm at 14 mmHg and from 6.2 mm to 11.9 mm at 7 mmHg. Our proposed registration method achieved mean TRE values below 5 mm across all these samples despite featuring large deformations, with the exception of one sample where the intraoperative surface was highly compromised due to a highly inaccurate stereo reconstruction. Hence, the widely adopted linear elastic biomechanical approach can yield reasonable registration accuracy, even within these typically large deformation ranges. This observation is also shared by previous works \cite{rucker2013mechanics, heiselman2024image, wittek2009unimportance} that also suggested that a linear elastic model is perhaps adequate to achieve sufficiently accurate non-rigid organ registration.}

\textcolor{black}{The liver can also be represented using a nonlinear biomechanical model. While nonlinear models may be considered superior to linear elastic models at capturing large deformations, the material properties required for such models to be sufficiently realistic and accurate vary significantly among patients and are difficult to measure, especially during surgery. Additionally, using a nonlinear model increases computational complexity, as the tangent stiffness matrix must be recalculated with each algorithm iteration. To incorporate a nonlinear model into our method, line 5 of Algorithm 1 can be modified to utilize the FEM for calculating deformations and the tangent stiffness matrix. Various nonlinear models can be selected for this purpose. For further details, we refer the reader to the works described in \cite{wittek2009unimportance, zhang2017deformable, peterlik2018fast, plantefeve2016patient}, leaving this aspect for future exploration.}



\subsection{Potential Challenges}

\textcolor{black}{The intraoperative point cloud, collected using an optically tracked stylus, is very sparse compared to other imaging modalities. However, despite this sparsity, the proposed method still achieved accurate results. To address extremely sparse intraoperative point clouds, a grid-fitting interpolation method similar to the one used by Collins \textit{et al.} \cite{collins2017improving} could be employed to recover denser surface points. However, it is recommended to avoid such sparse conditions during data collection, and hence, an extremely sparse data analysis would become essentially unrealistic.}

Before performing non-rigid registration, it is essential to (1) minimize intraoperative liver surface reconstruction errors, (2) ensure a high-quality surface with sufficient convergence, and (3) achieve accurate rigid alignment. These steps are critical prerequisites; otherwise, our method, like others, may yield inaccurate results. \textcolor{black}{Integrating real-time imaging modalities, such as ultrasound or cone-beam CT, could potentially relax these requirements by providing additional constraints, such as vascular features (i.e., ligaments visible in ultrasound) or a complete intraoperative surface from cone-beam CT. }








\section{Conclusion}

We have presented a \textcolor{black}{unique} non-rigid 3D-3D registration method that seamlessly integrates a biomechanical model into the surface-matching term without relying on priors. By embedding the biomechanical model in the surface-matching term, the estimated deformation is forced to preserve the geometry of the preoperative model \textcolor{black}{without an additional strain energy term}. Experiments on our \textit{in silico} phantom, \textit{in vitro} phantom datasets, and two publicly available datasets demonstrate the method's efficacy in achieving accurate registration results and preserving original geometry. Furthermore, our constructed phantom dataset and registration method represent significant contributions to liver registration benchmarking and, hence, valuable additions considering the limited availability of open-source datasets and methods.

\textcolor{black}{As part of our future work, we intend to research the effects of nonlinear hyperelastic materials, incorporating learning-based methods, automating the process of determining the stabilizing parameter and the potential combination of this proposed method with other real-time intraoperative imaging modalities. We also note that the 3D-3D non-rigid registration problem tackled here shares many similarities with those used for other organs, such as the prostate, kidney, brain, and breast, and hence, we would be interested in investigating the performance of the proposed method on other organs.}





\bibliographystyle{ieeetr}
\bibliography{refs}

\end{document}